# Electrochemical Modeling of Calendar Capacity Loss of Nickel-Manganese-Cobalt (NMC)-Graphite Lithium Ion Batteries

Boman Su[#], Xinyou Ke[#] and Chris Yuan[*]

Department of Mechanical and Aerospace Engineering, Case Western Reserve University,

Cleveland, Ohio 44106, United States

[#]These authors contributed equally; [*]Corresponding author: chris.yuan@case.edu

## Abstract

Li-ion batteries with nickel-manganese-cobalt (NMC) cathode and graphite anode are popularly used in portable electronic devices and electric vehicles. Calendar loss of the lithium ion battery is a dominating factor in battery degradation. However, few modeling work was reported on studying the calendar capacity loss of NMC-graphite Li-ion batteries. In this work, an electrochemical model for NMC-graphite Li-ion battery was developed to investigate its calendar loss behavior. Various factors affecting the calendar loss of the NMC-graphite batteries were systematically investigated, with the results validated with experimental data of a Sanyo 18,650 cylindrical cell. It was found that at 25 ºC working temperature and 100% SOC, the capacity drops 6.4% of its original capacity after 10 months. Also, when the anode particle size decreases from 26.2 μm to 6.55 μm, the capacity drop ratio is over 22% after 10 months under the same operation condition. Our simulation results demonstrate that a smaller SOC, a lower cell working temperature and a larger particle size could prolong the battery life during the storage period. This modeling work can help better understand the calendar loss behavior of NMC-graphite Li-ion batteries, and serve as a robust reference for the battery performance optimization in future.

***Keywords:*** Electrochemical modeling; battery degradation; capacity degradation; calendar loss; NMC-graphite Li-ion batteries.

**Nomenclature**

| | |
|---|---|
| $a$ | specific surface area of the electrode (m$^{-1}$) |
| $c$ | concentration (mol L$^{-1}$) |
| $D$ | diffusion coefficient (m$^2$ s$^{-1}$) |
| $d$ | firm thickness (μm) |
| $E$ | energy (J mol$^{-1}$) |
| $F$ | Faraday constant (96,485 C mol$^{-1}$) |
| $i$ | current density (mA cm$^{-2}$) |
| $k$ | reaction rate (m s$^{-1}$) |
| $L$ | thickness (μm) |
| $n$ | number of transferred lithium ions |
| $OCV$ | open-circuit voltage (V) |
| $R$ | universal gas constant (8.3143 J mol$^{-1}$ K$^{-1}$) |
| $R_{SEI}$ | resistance of solid-electrolyte interphase (Ω m$^{-2}$) |
| $r$ | radius of electrode particle (m) |
| $SOC$ | state of charge |
| $T$ | cell temperature (K) |
| $t+$ | transport number |
| $V$ | molar volume (m$^3$ mol$^{-1}$) |
| $x$ | stoichiometric number in the anode |
| $y$ | stoichiometric number in the cathode |

*Greek symbols*

| | |
|---|---|
| $\alpha$ | transfer coefficient |
| $\varepsilon$ | volume fraction |
| $\eta$ | over-potential of electrode reaction (V) |
| $\kappa$ | ionic conductivity (mS cm$^{-1}$) |
| $\sigma$ | electronic conductivity (mS cm$^{-1}$) |
| $\phi$ | potential (V) |

*Subscripts*

| | |
|---|---|
| $a$ | anodic |
| $act$ | activation |
| $app$ | applied value |
| $c$ | catholic |
| $cell$ | battery cell |
| $eq$ | equilibrium |
| $in$ | intercalation |
| $iso$ | isolation |
| $l$ | liquid phase |
| $max$ | maximum |
| $min$ | minimum |
| $neg$ | negative electrode |
| *pos* | positive electrode |
| *r* | radial direction in electrode particle |
| *ref* | reference value |
| *sep* | separator |
| *s* | solid phase |
| *side* | side reaction |
| *SEI* | solid-electrolyte interphase |
| *0* | initial |
| | |
| *Superscripts* | |
| *eff* | effective |

# 1. Introduction

Over the past decades, lithium ion batteries (LIBs) have been widely adopted in various portable electronic devices and electric vehicles, owing to their relatively high energy and power densities [1-4]. Nowadays the most popular LIBs is $Li_x[MnNiCo]O_2$ (NMC)-graphite lithium ion battery [5]. According to a report on estimating U.S. lithium ion battery raw material markets, NMC, together with LCO ($LiCoO_2$), LFP ($LiFePO_4$), LMO ($LiMn_2O_4$) and NCA ($LiNi_{0.8}Co_{0.15}Al_{0.05}O_2$) are the five most demanding cathode materials during the next five years [6]. Among these cathode materials, NMC has been favored by many manufacturers due to its various merits. On one hand, NMC can last longer for a single charge due to its high energy density compared with other cathode material except for NCA. On the other hand, NCA is prone to be less thermal stable compared with NMC [7]. However, battery capacity loss is a critical issue for NMC-graphite LIBs since the aging of NMC cathode exhibit high sensitivity to the temperature [7-10] and cell voltage [11-13].

Generally, the capacity loss of lithium ion battery can be divided into two categories: cycling loss and calendar loss. Cycling and calendar losses mean the irreversible capacity loss during the operation period and storage period, respectively. Studies show that in practical applications, such as EVs, calendar loss is more significant than cycling loss since the storage period is much longer than the operation period [14]. The main cause of the calendar loss was considered from a slow "self-discharge" process [15,16]. This process is a combined effect of several factors, such as SEI formation [17,18], electrode corrosion [15], partial dissolution of active materials [15]. For example, Redondo et al. found that the capacity loss of self-discharge of NMC-graphite battery varies from around 2.3% per month for a fresh cell to 0.3% per month after 400 days [19]. The capacity loss of self-discharge in Li-ion cells can be classified to be reversible or irreversible [20].

The reversible portion can be recovered by recharging the battery, while the irreversible portion is a permanent capacity loss considered as the most significant portion of the calendar loss [18,21,22]. The portion of the reversible or irreversible capacity loss depends on the conditions of the tested cell, such as temperature, SOC and humidity [17,23]. To date, studies on calendar aging behaviors of NMC-graphite LIBs are insufficient, and thus it is urgent to fully understand their capacity degradation behavior and further help to optimize the battery performance.

To study the capacity degradation of NMC-based LIBs, modeling has been considered as an effective approach to achieve this goal. Both physics-based and semi-empirical models of NMC-based LIBs have been reported in the literature. For example, some semi-empirical models were developed to estimate both the calendar and cycling capacity loss of NMC-based LIBs [24-27]. They used experimental datasets to fit the empirical functions directly, i.e., kinetics equation [27], general exponential function [25] and Arrhenius equation [24,26,27] by treating the operating conditions as variables. Laresgoiti et al. developed an electro-mechanical model for an NMC Li-ion battery to capture the cycling loss due to the mechanical degradation of the solid electrolyte interphase (SEI) layer [28]. They suggested that the battery aging can be greatly mitigated at an intermediate state of charge (SOC) and a relatively small depth of discharge (DOD). Fu et al. developed a model of an NMC pouch cell, and they considered the electrodeposition and losses of active material and electrolyte in the model [29]. Their work shows that a high SOC and discharge rate could accelerate battery cycling loss due to the promotion side reactions. These recent findings provide insightful understandings of NMC-based Li-ion battery degradation and the basis for optimizing the battery performance.

To date, most of the models developed for estimating the calendar loss of NMC-based LIBs during the storage period are semi-empirical [30-33]. For example, a square root time dependency function based on the SEI formation theory was commonly used [32]. A nearly linear trend of the capacity degradation curve was observed if the square root time dependency function was used [8,30]. To better capture the calendar aging behavior, a variety of "in-between" functions were developed. For example, Schmalstieg et al. used a superposition of linear and square root functions, together with a power function [31]. Also, a generalized power function was used to treat the exponent as a fitting parameter [30]. Moreover, Ecker et al. applied a superposition of a logarithmic function and a square root function [33]. Although these models can make a relatively rough prediction for the calendar loss of LIBs, they all contain some constructs that are somehow arbitrary without any solid physics foundations. Thus, they potentially lack accuracy when applying to a different type of battery chemistries or a wide range of storage conditions. To overcome the disadvantages of these semi-empirical models, attempts on developing physics-based models of NMC-based Li-ion batteries to estimate the calendar loss have been reported. For example, Lu et al. developed a degradation model considering several degradation factors, such as loss of active material [34]. However, this model did not consider the mass balance of the battery system, and it used an empirical function to describe the SEI layer formation controlled by the diffusion. Jana et al. developed a single particle model considering the SEI formation and electrolyte oxidation at the cathode [35]. However, this model ignores the losses of electrolyte and active materials at the anode during calendar aging. Thus, a more accurate physics-based modeling analysis of capacity calendar loss for NMC-graphite Li-ion batteries is desired.

In this work, a physics-based model was developed to investigate the calendar capacity loss behavior of an NMC-graphite Li-ion battery. Various factors, such as SOC, cell working temperature and particle size, on calendar loss were quantitatively examined. Modeling results were finally validated with experimental data from literature. It is expected that this study can provide certain useful guidance on how to minimize the calendar loss for Li-ion batteries during the storage condition.

## 2. Modeling description

A pseudo-2D model of a lithium ion battery with a graphite anode, a separator, and a Li$_x$[MnNiCo]O$_2$ (NMC) cathode was developed and shown in Figure 1. During discharge, Li$^+$ stored in the anode particles de-intercalates and move to the cathode through the electrolyte and separator. Simultaneously, electrons generated will be transferred through the external circuit. This pseudo-2D model was simplified for a 3D commercial 18,650 cylindrical NMC-graphite Li-ion battery cell with a capacity of 2.05 Ah simulated in this work [36]. The transport along the height direction of the 3D cylindrical cell was assumed to be uniform [37].

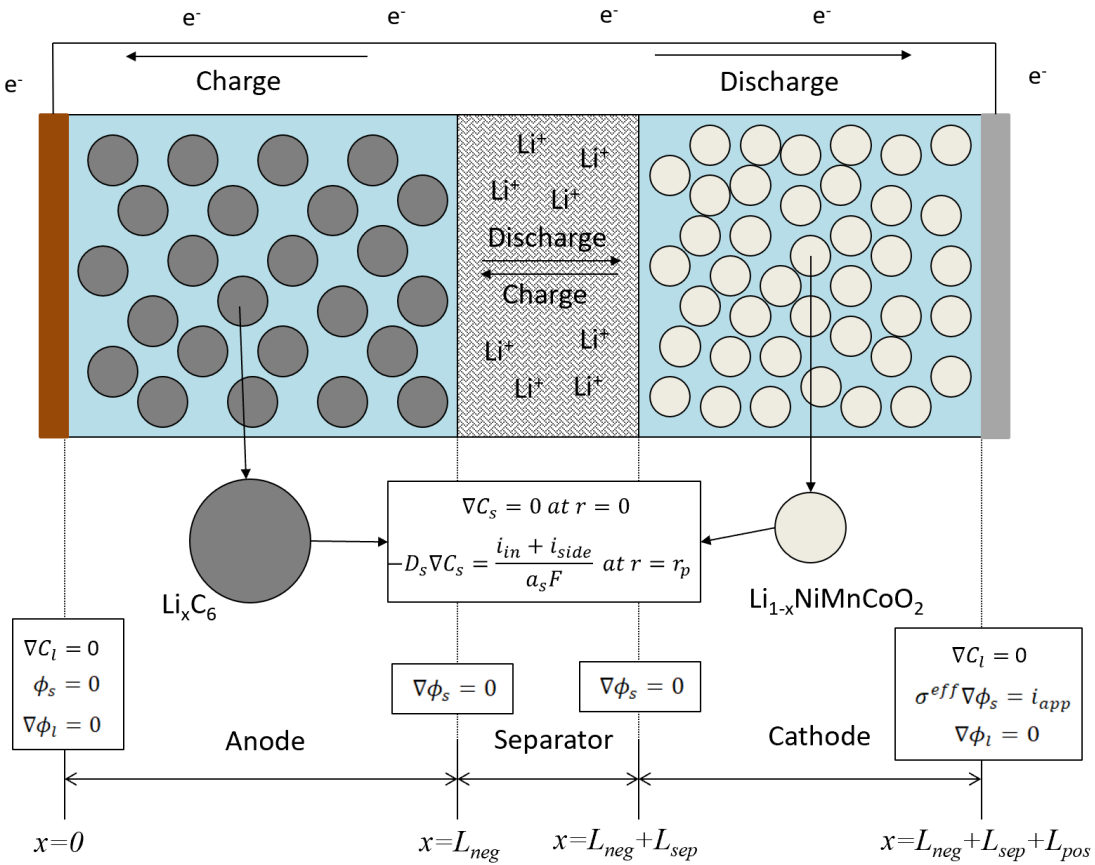

**Figure 1: A pseudo-2D model of NMC-graphite lithium ion battery.**

## 2.1 Electro-kinetics and side reactions

Li-ion intercalation and de-intercalation occur at both the anode and cathode during the charge/discharge. Apart from the intercalation and de-intercalation, two major side reactions also take place at the surfaces of anode particles [29]:

$$2Li^+ + 2e^- + EC == CH2 + Li_2CO_3 \downarrow \quad (1)$$

$$2Li^+ + 2e^- + 2EC == CH2 + (CH_2OCO_2Li)_2 \downarrow \quad (2)$$

Thus, the electrode kinetic is considered as a combination of the intercalation reaction and the side reaction [38]:

$$i_{total} = i_{in} + i_{side} \quad (3)$$

Where, $i_{in}$ is the intercalation current density and $i_{side}$ is the side reaction current density. The intercalation process can be described using Butler-Volmer equation [38,39]:

$$i_{in} = k_i(C_s)^{\alpha_c}\left(C_l(C_{s,max} - C_s)\right)^{\alpha_a}\left(e^{\frac{\alpha_a nF}{RT}\eta} - e^{-\frac{\alpha_c nF}{RT}\eta}\right) \quad (4)$$

Where, $k_i$ ($i$ = pos or neg) is the reaction rate constant, $C_s$ is the solid phase Li$^+$ concentration at the particle surface, $C_l$ is the concentration of the liquid phase Li$^+$ in the electrolyte, $C_{s,max}$ is the maximum solid-phase Li$^+$ concentration, $\alpha_a/\alpha_c$ is the charge transfer coefficient of the anodic/cathodic reaction, $n$ is the number of electrons transferred, $F$ is the Faraday's constant, $R$ is the universal gas constant, $T$ is the battery operation temperature, and $\eta$ is the over-potential of the intercalation given by [37,40]:

$$\eta = \phi_s - \phi_l - \phi_{eq} - R_{SEI}(i_{in} + i_{side}) \tag{5}$$

Where, $\phi_s$ is the potential of the solid phase electrode, $\phi_l$ is the potential of the liquid phase electrolyte, $E_{eq}$ is the equilibrium potential of the solid phase electrode, and $R_{SEI}$ is the resistance of the SEI layer. Degradation under three main side reactions was modeled in this work: (1) growth of SEI layer, (2) loss of the active material, and (3) loss of the electrolyte. The SEI layer might be formed due to the side reactions (1) and (2). Assuming that $i_{side}$ is not a function of dimension, $R_{SEI}$ can be calculated as [29,41]:

$$R_{SEI} = \frac{d_{SEI,0}}{\kappa_{SEI}} + \frac{V_{SEI}}{2F\kappa_{SEI}} \int_0^{t_{st}} i_{side} dt \tag{6}$$

Where, $d_{SEI,0}$ is the initial thickness of SEI layer, $\kappa_{SEI}$ is the ionic conductivity of SEI layer, $V_{SEI}$ is the molar volume of SEI layer, $t_{st}$ is the battery storage time. The SEI layer can electrically isolate some graphite particles, and thus these particles cannot participate in the Li$^+$ intercalation and de-intercalation. Consqunetly, the growth of SEI layer also causes the loss of the active materials. The loss of active material was described using the change of the volume fraction of the electrode, and it was given as [29]:

$$\Delta\varepsilon_s = -k_{iso} \frac{V_{SEI}}{2F} \int_0^{t_{st}} i_{side} a_s dt \tag{7}$$

Where, $k_{iso}$ is a dimensionless coefficient that describes the rate of the active material isolating from the chemical reaction, and $a_s$ is the specific area of the active material, which can be calculated as [37,42]:

$$a_s = \frac{3\varepsilon_s}{r_p} \tag{8}$$

The loss of electrolyte can be described using the change of the volume fraction of the electrolyte [29]:

$$\Delta\varepsilon_l = -\frac{\alpha V_e}{F}\int_0^{t_{st}} i_{side}a_s dt \tag{9}$$

Where, $\alpha$ is a coefficient that indicates how many moles of electrolyte can be consumed when one mole of Li$^+$ is consumed, and $V_e$ is the molar volume of the electrolyte. The formula for governing the side reaction current can be given [38]:

$$i_{side} = i_{0,side}\left(e^{\frac{\alpha_{a,side}n_{side}F}{RT}\eta_{side}} - e^{\frac{-\alpha_{c,side}n_{side}F}{RT}\eta_{side}}\right) \tag{10}$$

Where, $i_{0,side}$ is the exchange current density of the side reaction, $\alpha_{a,side}/\alpha_{c,side}$ is the charge transfer coefficient of the anodic/cathodic side reaction, $n_{side}$ is the number of ions involved in the side reaction, $\eta_{side}$ is the overpotential of the side reaction, and it yields [37,40]:

$$\eta_{side} = \phi_s - \phi_l - \phi_{eq,side} - R_{SEI}(i_{in} + i_{side}) \tag{11}$$

Where, $\phi_{eq,side}$ is the equilibrium potential of the side reaction, and it is affected by the concentration of the liquid phase Li$^+$ in the electrolyte [29]:

$$\phi_{eq,side} = \phi_{eq,side,0} + \frac{RT}{n_{side}F} \ln\left(\frac{c_l}{c_{l,0}}\right) \tag{12}$$

Where, $E_{eq,side,0}$ is the equilibrium reaction for the lithium/solvent reduction reaction when $c_l$ is equal to $c_{l,0}$, $c_{l,0}$ is the initial concentration of the electrolyte. The mass balance of the electrode yields [41,43]:

$$\frac{\partial C_s}{\partial r} = \frac{D_s}{r^2} \frac{\partial}{\partial r}\left(r^2 \frac{\partial C_s}{\partial r}\right) \tag{13}$$

Where, $D_s$ is the solid Li$^+$ diffusion coefficient, and $r$ is the direction that is normal to the surface of the particle. The mass balance of the electrolyte yields [41,43]:

$$\varepsilon_l \frac{\partial C_l}{\partial t} = \frac{\partial}{\partial x}\left(D\varepsilon_l^{1.5} \frac{\partial C_l}{\partial x}\right) + \left(\frac{1-t_+}{F}\right)(i_{in} + i_s) \tag{14}$$

Where, $\varepsilon_l$ is the liquid phase volume fraction, $D$ is the diffusion coefficient of the electrolyte and $t_+$ is the transport number of the electrolyte. The charge balance of the electrode yields [40,44]:

$$\sigma\varepsilon_s^{1.5} \frac{\partial^2 \phi_s}{\partial x^2} = i_{in} + i_s \tag{15}$$

Where, $\sigma$ is the solid phase conductivity and $\varepsilon_s$ is the solid phase volume fraction. The charge balance of electrolyte yields [40,44]:

$$\frac{\partial}{\partial x}\left(\kappa \varepsilon_l^{1.5} \frac{\partial \phi_l}{\partial x}\right) + \frac{\partial}{\partial x}\left(\kappa_D \varepsilon_l^{1.5} \frac{\partial \ln C_l}{\partial x}\right) + (i_{in} + i_s) = 0 \qquad (16)$$

Where, $\kappa$ is the liquid phase conductivity, $\varepsilon_l$ is the liquid phase volume fraction, and $\kappa_D$ is the diffusion conductivity given by [43,44]:

$$\kappa_D = \frac{2RT\kappa}{F}(1 - t^+) \qquad (17)$$

All the above-mentioned equations were coupled to solve for four unknown parameters: $\phi_s$, $\phi_l$, $C_s$ and $C_l$. Simultaneously, $R_{SEI}$, $\Delta \varepsilon_l$ and $\Delta \varepsilon_s$ were also solved.

## 2.2 Assumptions and boundary conditions

Several assumptions were made for this model: (1) this pseudo-2D model was simplified for the 3D NMC-graphite Li-ion battery because the electrochemical reaction along the height direction (the 3rd dimension) of the Li-ion battery could be uniform [37,45]; (2) the irreversible calendar loss was assumed to attribute to the "self-discharging" process [18,21]; (3) no major side reactions were considered at the cathode side [29] and (4) the calendar aging process was treated as isothermal since the thermal effect could be neglected during the battery "self-discharge" process under a very low discharge rate. For the mass balance of the electrode, the diffusion occurs at the boundary of the active material particle, and boundary conditions were given [43,44,46]:

$$\frac{\partial C_s}{\partial r} = 0 \; at \; r = 0 \tag{18}$$

$$-D_s \frac{\partial C_s}{\partial r} = \frac{i_{in} + i_s}{a_s F} \; at \; r = r_p \tag{19}$$

Where, $r_p$ is the particle radius of the active material. For the mass balance of the electrolyte, no flux of lithium ion was defined at the boundaries:

$$\frac{\partial C_l}{\partial x} = 0 \; at \; x = 0 \; and \; x = L_{batt} \tag{20}$$

For the charge balance of the electrode, the anode was ground:

$$\phi_s = 0 \; at \; x = 0 \tag{21}$$

The current was applied on the cathode current collector:

$$\sigma \varepsilon_s^{1.5} \frac{\partial \phi_s}{\partial x} = 0 \; at \; x = 0, x = L_{neg} \; and \; x = L_{neg} + L_{sep} \tag{22}$$

$$\sigma \varepsilon_s^{1.5} \frac{\partial \phi_s}{\partial x} = i_{app} \; and \; x = L_{neg} + L_{sep} + L_{pos} \tag{23}$$

For the charge balance of the electrolyte, "zero" ionic current was defined:

$$\frac{\partial \phi_l}{\partial x} = 0 \text{ at } x = 0 \text{ and } x = L_{neg} + L_{sep} + L_{pos} \tag{24}$$

## 2.3 Modeling parameters and computational details

Parameters used in this electrochemical model are listed in Table 1. These parameters mainly include physical properties of electrode and electrolyte materials, side reaction factors, operation conditions and cell dimensions. Some physical properties are related to the stoichiometry of chemical reactions or concentrations of the active species as shown in Figure 2. To better capture each electrode voltage curve, effects of both the stoichiometry and temperature on the open circuit potential is considered in this model. Figure 2 (a) and (b) shows a relationship between the open circuit potentials and the stoichiometry at the anode and cathode, respectively. It can be seen that the open circuit potential at both the anode and cathode decreases with the stoichiometry $x$ or $y$. Figure 2 (c) and (d) indicate the function of the temperature derivative of the open circuit potential vs. the stoichiometry at both electrodes. Figure 2 (e) and 2 (f) describes the corresponding ionic conductivity and diffusion coefficient of electrolyte with respect to its concentration. It can be seen that the Li$^+$ diffusion coefficient is within the range of $1.4 \times 10^{-10}$ to $3.6 \times 10^{-10}$ m$^2$ s$^{-1}$, and the ionic conductivity is between 0 and 8 mS cm$^{-1}$ for the electrolyte at 25 °C, depending on the concentration [47]. Based on the experimental data from the literature, these properties were fitted as the polynomials with the concentration or stoichiometry variables. Side reaction factors include the side reaction current density, the charge transfer coefficient of the side reaction, and the active material isolated rate. The side reaction current density was estimated to be $1.1 \times 10^{-6}$ A m$^{-2}$ at 25 °C based on the calendar loss experimental data [36]. Charge transfer coefficients on the cathodic and anodic sides were estimated to be 0.7 and 0.3 [29], respectively. Thermal related parameters as shown in Table 1 use the Arrhenius equation to describe their temperature dependence [48]:

$$\psi = \psi_{ref} \exp\left[\frac{E_{act}}{R}\left(\frac{1}{T_{ref}} - \frac{1}{T}\right)\right] \tag{25}$$

Where, $\psi$ could be referred to the diffusion coefficient of Li$^+$ in the solid active material and electrolyte, the ionic conductivity of the electrolyte, the transport number and the exchange current density of side reactions, $E_{act}$ is the active energy of the variable $\psi$, and $\psi_{ref}$ represents the value of $\psi$ at $T_{ref} = 25°C$.

**Table 1: Physical and dimensional properties of NMC-graphite Li-ion batteries.**

| Symbol | Value/Expression | Unit | Description | Reference |
|---|---|---|---|---|
| $\alpha$ | 0.75 | 1 | Moles of electrolyte consumed when 1 mol of Li$^+$ is consumed | [29] |
| $\alpha_{a,side}$ | 1 $-\alpha_{c,side}$ | 1 | Anodic charge transfer coefficient of the side reaction | [29] |
| $\alpha_a$, $\alpha_c$ | 0.5 | 1 | Anodic/Cathodic charge transfer coefficient | [29] |
| $\alpha_{c,side}$ | 0.7 | 1 | Catholic charge transfer coefficient of the side reaction | [29] |
| $\phi_{eq,neg}$ | $0.1493 + 0.8493\,exp(-61.79x) + 0.3824\,exp(-665.8x) - exp(39.24x - 41.92) - 0.03131\,atan(25.59x - 4.099) - 0.009434\,atan(32.49x - 15.74)$ (x=negative material stoichiometry) | V | Negative electrode equilibrium potential | [37] |
| $\phi_{eq,pos}$ | $-2.5947y^3 + 7.1062y^2 - 6.9922y + 6.0826 - 0.000054549\,exp(124.23y - 114.2593)$ (y=positive material stoichiometry) | V | Positive electrode equilibrium potential | [37] |
| $\phi_{eq,side,0}$ | 0.21 | V | Equilibrium potential of the side reaction | [29] |
| $\sigma_{neg}$ | 1000 | mS cm$^{-1}$ | Negative electrode solid phase conductivity | [49] |
| $\sigma_{pos}$ | $T = 25$: $133.2y^2 + 73.2y + 1.1$ $T = 50$: $264y^3 - 197y^2 + 185y - 0.5$ (y=positive material stoichiometry) | mS cm$^{-1}$ | Positive electrode solid phase conductivity | [50] |
| $\kappa_l$ | $1.147x^3 - 22.38x^{1.5} + 29.15x$ (x=cl in mol L$^{-1}$) | mS cm$^{-1}$ | Liquid phase ionic conductivity | [47] |
| $\kappa_{SEI}$ | $4.2 \times 10^{-5}$ | mS cm$^{-1}$ | SEI layer ionic conductivity | [51] |
| $\varepsilon_{l,neg}$ | 0.26 | 1 | Negative electrode liquid phase volume fraction | [52] |
| $\varepsilon_{l,pos}$ | 0.37 | 1 | Positive electrode liquid phase volume fraction | [52] |

| Symbol | Value | Unit | Description | Ref. |
|---|---|---|---|---|
| $\varepsilon_{s,neg}$ | 0.58 | 1 | Negative electrode solid phase volume fraction | [29] |
| $\varepsilon_{s,pos}$ | 0.5 | 1 | Positive electrode solid phase volume fraction | [29] |
| $D_l$ | $7.588 \times 10^{-11} c_l^2 - 3.036 \times 10^{-10} c_l + 3.654 \times 10^{-10}$ ($c_l$ in mol L$^{-1}$) | m$^2$ s$^{-1}$ | Liquid phase lithium ion diffusion coefficient | [47] |
| $D_{s,neg}$ | $1.55 \times 10^{-14}$ | m$^2$ s$^{-1}$ | Negative electrode solid phase diffusion coefficient | [53] |
| $D_{s,pos}$ | $1.904 \times 10^{-14}\exp(-7.873y) + 3.164 \times 10^{-14}\exp(-2.064y)$ (y=positive material stoichiometry) | m$^2$ s$^{-1}$ | Positive electrode solid phase diffusion coefficient | [50] |
| $E_{act,Dl}$ | 12,360 | J mol$^{-1}$ | Activation energy of the liquid phase diffusion coefficient | [54] |
| $E_{act,t+}$ | 9,890 | J mol$^{-1}$ | Activation energy of the transport number | [54] |
| $E_{act,Ds,neg}$ | 20,000 | J mol$^{-1}$ | Activation energy of the negative electrode solid phase diffusion coefficient | [55] |
| $E_{act,Ds,pos}$ | 93,533 | J mol$^{-1}$ | Activation energy of the positive electrode solid phase diffusion coefficient | Fitted by experimental data from [50] |
| $E_{act,SEI}$ | 65,000 | J mol$^{-1}$ | Activation energy of the SEI formation | Assumed base on 50°C calendar loss data |
| $E_{act,\kappa}$ | 15,840 | J mol$^{-1}$ | Activation energy of the electrolyte ionic conductivity | [54] |
| $L_{neg}$ | $4 \times 10^{-5}$ | m | Length of the negative electrode | [52] |
| $L_{pos}$ | $3.5 \times 10^{-5}$ | m | Length of the separator | [52] |
| $L_{sep}$ | $2 \times 10^{-5}$ | m | Length of the positive electrode | [52] |
| $V_e$ | 56.8 | cm$^3$ mol$^{-1}$ | Molar volume of the electrolyte | [29] |
| $V_{SEI}$ | 2 | cm$^3$ mol$^{-1}$ | Molar volume of the SEI layer | [29] |
| $c_{l,0}$ | 1,000 | mol m$^{-3}$ | Initial electrolyte salt concentration | [52] |
| $c_{s,0,neg}$ | $0.936 c_{s,max,neg}$ | mol m$^{-3}$ | Initial negative state of charge | Assumed based on the voltage curve |

| Symbol | Value | Units | Description | Ref |
|---|---|---|---|---|
| $c_{s,0,pos}$ | $0.442 c_{s,max,pos}$ | mol m$^{-3}$ | Initial positive state of charge | [29] |
| $c_{s,max,neg}$ | 31,000 | mol m$^{-3}$ | Negative maximum solid phase concentration | [48] |
| $c_{s,max,pos}$ | 48,500 | mol m$^{-3}$ | Positive maximum solid phase concentration | Assumed based on voltage curve |
| $\dfrac{d\phi_{eq,neg}}{dT}$ | $-58.294x^6 + 189.93x^5 - 240.4x^4 + 144.32x^3 - 38.87x^2 + 2.8642x + 0.1079$ (x=negative material stoichiometry) | V K$^{-1}$ | Negative electrode temperature derivative of the equilibrium potential | [37] |
| $\dfrac{d\phi_{eq,pos}}{dT}$ | $-190.34y^6 + 733.46y^5 - 1172.6y^4 + 995.88y^3 - 474.04y^2 + 119.72y - 12.457$ (y=positive material stoichiometry) | V K$^{-1}$ | Positive electrode temperature derivative of the equilibrium potential | [37] |
| $d_{SEI,0}$ | $2 \times 10^{-9}$ | m | Initial SEI layer thickness | [29] |
| $i_{0,side}$ | $1.1 \times 10^{-6}$ | A m$^2$ | Exchange current density of the side reaction | Assumed based on 25°C calendar loss data |
| $i_{1C}$ | $0.5 F c_{s,max,pos} \varepsilon_{s,pos} L_{pos} / 3600$ | A m$^2$ | 1C discharge current | [37] |
| $k_{iso}$ | 27.3 | 1 | Isolation rate of the active anode material due to the SEI layer | [29] |
| $k_{neg}$ | $1.55 \times 10^{-11}$ | m s$^{-1}$ | Negative reaction rate coefficient | [55] |
| $k_{pos}$ | $4.38 \times 10^{-11}$ | m s$^{-1}$ | Positive reaction rate coefficient | [41] |
| $n_{side}$ | 1 | 1 | Number of ions involved in the side reaction | [29] |
| $r_{p,neg}$ | $2.62 \times 10^{-5}$ | m | Negative electrode particle radius | [46] |
| $r_{p,pos}$ | $1.07 \times 10^{-5}$ | m | Positive electrode particle radius | [46] |
| $t_+$ | $-0.1291 c_l^3 + 0.3517 c_l^2 - 0.4893 c_l + 0.4287$ ($c_l$ in mol L$^{-1}$) | 1 | Transport number | [47] |

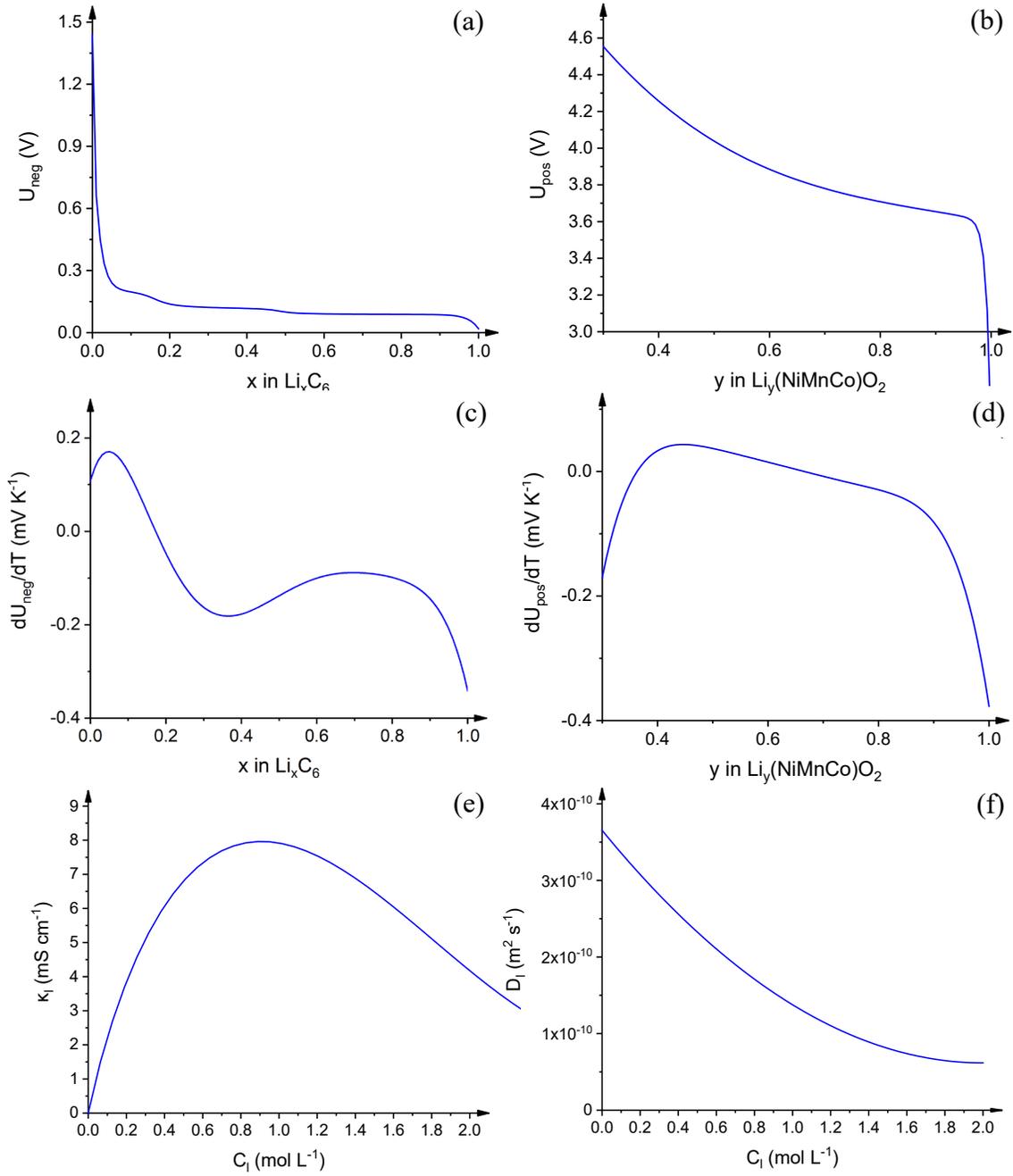

**Figure 2:** Relationships between some parameters with stoichiometry or concentrations used in the modeling of NMC-graphite Li-ion batteries through the polynomial fitting: (a) $U_{neg}$ vs. $x$ in $Li_xC_6$; (b) $U_{pos}$ vs. $y$ in $Li_y(NiMnCo)O_2$; (c) $dU_{neg}/dT$ vs. $x$ in $Li_xC_6$; (d) $dU_{pos}/dT$ vs. $y$ in $Li_y(NiMnCo)O_2$; (e) $k_l$ vs. $C_l$ and (f) $D_l$ vs $C_l$.

A total number of 420 unstructured rectangle elements were meshed for this pseudo-2D model, and the meshes at the interfaces between the anode/cathode and separator were refined. A mesh sensitivity analysis was also conducted to guarantee the accuracy of the simulation results. This pseudo-2D model was solved through the non-linear MUMPS solver in COMSOL Multiphysics 5.5 with a relative tolerance of $10^{-5}$. The simulation involves multiple modules, which include lithium ion battery interface, Global ODEs and DAEs, and Domain ODEs and DAEs.

## 3. Results and discussion

*3.1 charge-curve validation*

To validate the developed electrochemical model, the simulated charge curve was compared with the experimental data of a 2.05 Ah Sanyo UR18,650E NMC-graphite Li-ion battery cell reported in [36], as shown in Figure 3. For this simulation, following steps are used, as shown in Figure 4: (1) discharge at a constant current (CC) with $0.5i_{1C}$ ($i_{1C}$ is the current responding to the discharge rate of 1 C) until the voltage cutoff of 2.75 V; (2) discharge at a constant voltage (CV) until the current cutoff of $0.001i_{1C}$, and (3) charge at a constant current (CC) of $0.05i_{1C}$ until the voltage cutoff of 4.2 V. Step (1) and (2) can guarantee a state of charge (SOC) of 0% for the battery cell after this two-step discharge. It can be seen that the simulated charged curve of the cell voltage vs. capacity as shown in Figure 3 matches reasonably well with the experimental data. However, a slight discrepancy can be observed, and one major possible reason is that the maximum solid phase concentration might be underestimated in the cathode. Although enlarging this value from 48,500 mol m$^{-3}$ to 51,000 mol m$^{-3}$ could give an almost perfect fit for the voltage curve, but the trend of the capacity loss curve would be more linear. In fact, the maximum solid-state concentration of the NMC active material has a wide range of value in reported studies, from 36,100 mol m$^{-3}$ [41] to 87,593 mol m$^{-3}$ [46]. In this work, the assumed value of 48,500 mol m$^{-3}$ gives a reasonable prediction for both the voltage curve and the capacity loss curve.

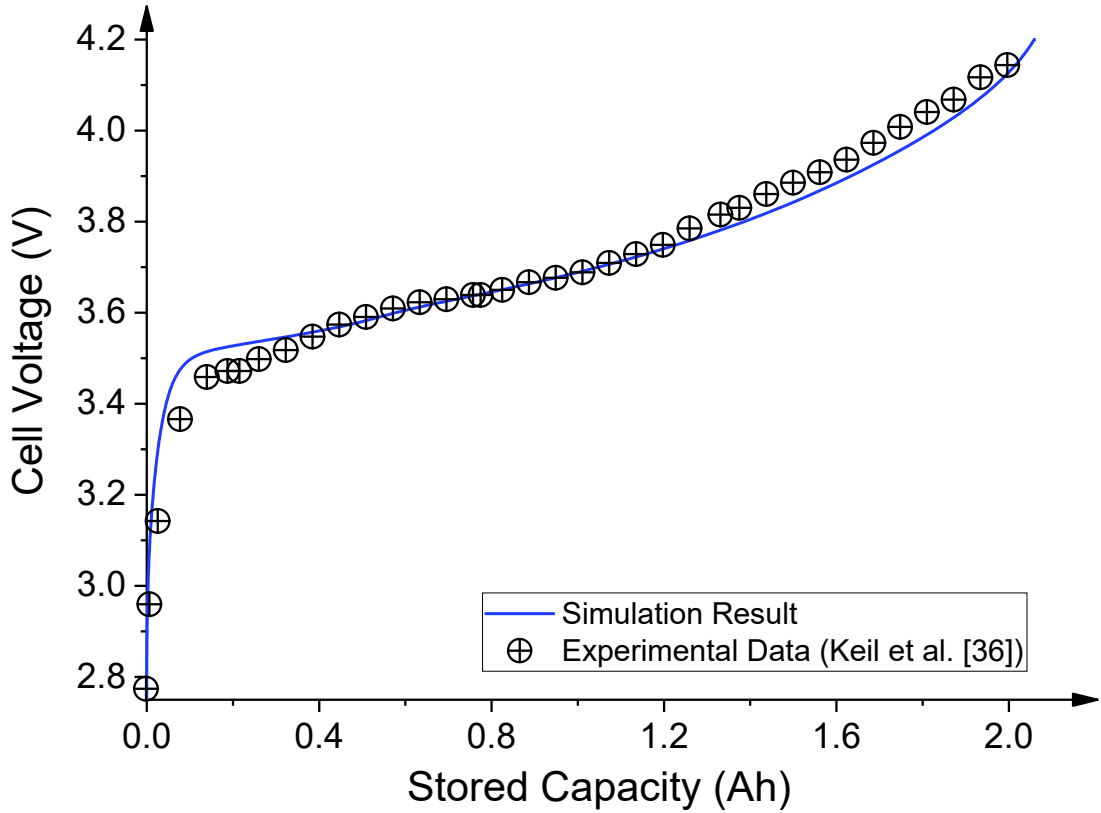

**Figure 3: A comparison on charge curve of cell voltage vs. capacity, for UR18650E: Simulation vs. experiment, and the experimental data is from Keil et al. [36].**

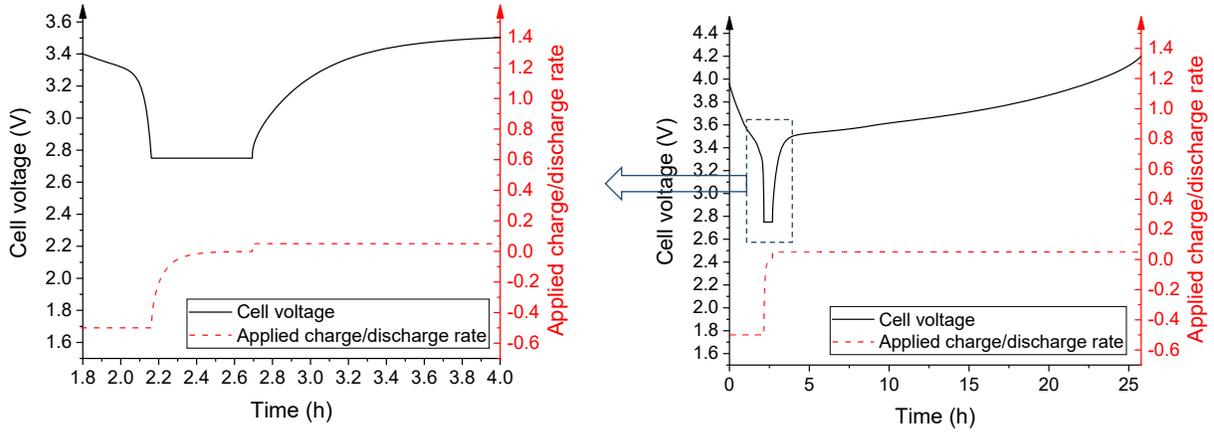

Figure 4: Simulation steps for charge-curve validation: constant current (cc)-constant voltage (cv)-constant current (cc).

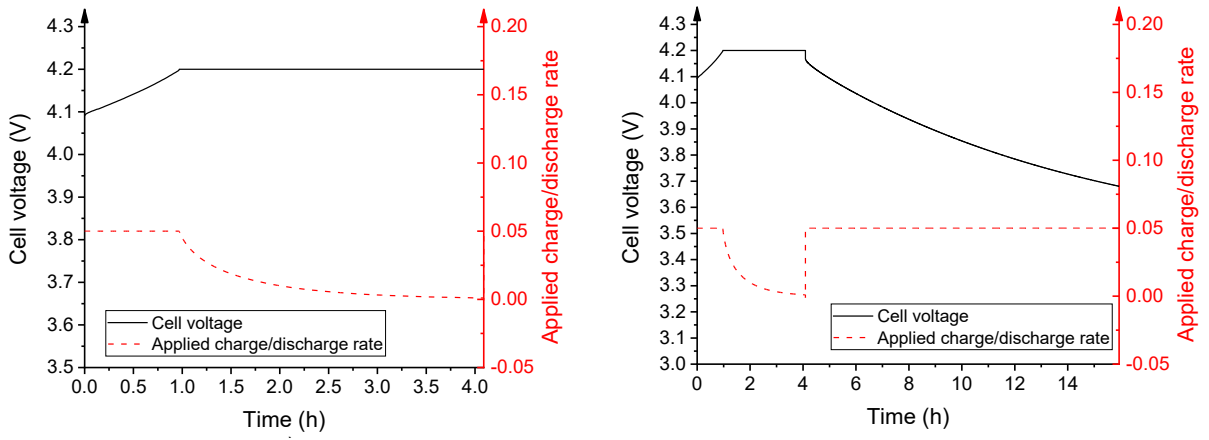

Figure 5: Simulation steps for making 100% and 50% state-of-charge (SOC) as the initial conditions for discharge.

*3.2 Effects of SOC, temperature and particle size on calendar loss*

The calendar loss of graphite-NMC Li-ion battery cell under the effects of various SOC, cell working temperature and electrode particle size has been studied. To validate the experimental data, simulations were conducted on two different SOCs (i.e. 50% and 100%), and two different temperatures (i.e. 25 °C and 50 °C). The particle size range of SFG44 graphite is from 8.5 μm to 48.4 μm [57]. A large range of particle size ratio (applied particle size/reference particle size) from 0.25 to 4 was studied.

For the case of 100% SOC, the following steps were used: (1) charge at a constant current (CC) of $0.05i_{1C}$ until the voltage cutoff of 4.2 V and (2) charge at a constant voltage (CV) of 4.2 V until the current cutoff of $0.001i_{1C}$ as shown in Figure 5 (a), and (3) discharge at a constant current (CC) of $10^{-5}i_{1C}$ for 10 months. Steps (1) and (2) initiate 100% SOC, and these steps provide a prerequisite for step (3) to simulate the calendar loss. For the case of 50% SOC, the following steps were used: (1) charge at a constant current (CC) of $0.05i_{1C}$ until the voltage cutoff of 4.2 V; (2) charge at a constant voltage (CV) of 4.2 V until the current cutoff of $0.001i_{1C}$; (3) discharge at a constant current (CC) of $0.05i_{1C}$ until the half of the maximum capacity as shown in Figure 5 (b), and (4) discharge at a constant current (CC) of $10^{-5}i_{1C}$ for 10 months. Steps (1), (2) and (3) initiate 50% SOC.

Figure 6 shows comparison of the simulated results and experimental results for the cases of 50% and 100% SOC under a constant cell working temperature of 25 °C and 50 °C. It can be seen that the capacity decays faster at the beginning and then gradually slower. Two possible reasons might explain this behavior: (1) the electrolyte tends to be more stable under a lower cell voltage [36]. Thus, as the cell voltage gradually decreases, the electrolyte oxidation would be mitigated. (2) As the thickness of the SEI layer continues to grow, it can more effectively prevent the electrolyte from further reduction [58]. At 25 °C, the capacity drops nearly 6.4% of its original capacity after 10 months under 100% SOC, while the capacity drops around 3.2% under 50% SOC. It was found that a higher SOC results in a larger permanent calendar loss. One possible reason is that a higher SOC leads to a graphite potential with a high absolute value [36], which can promote electrolyte reduction and SEI layer formation.

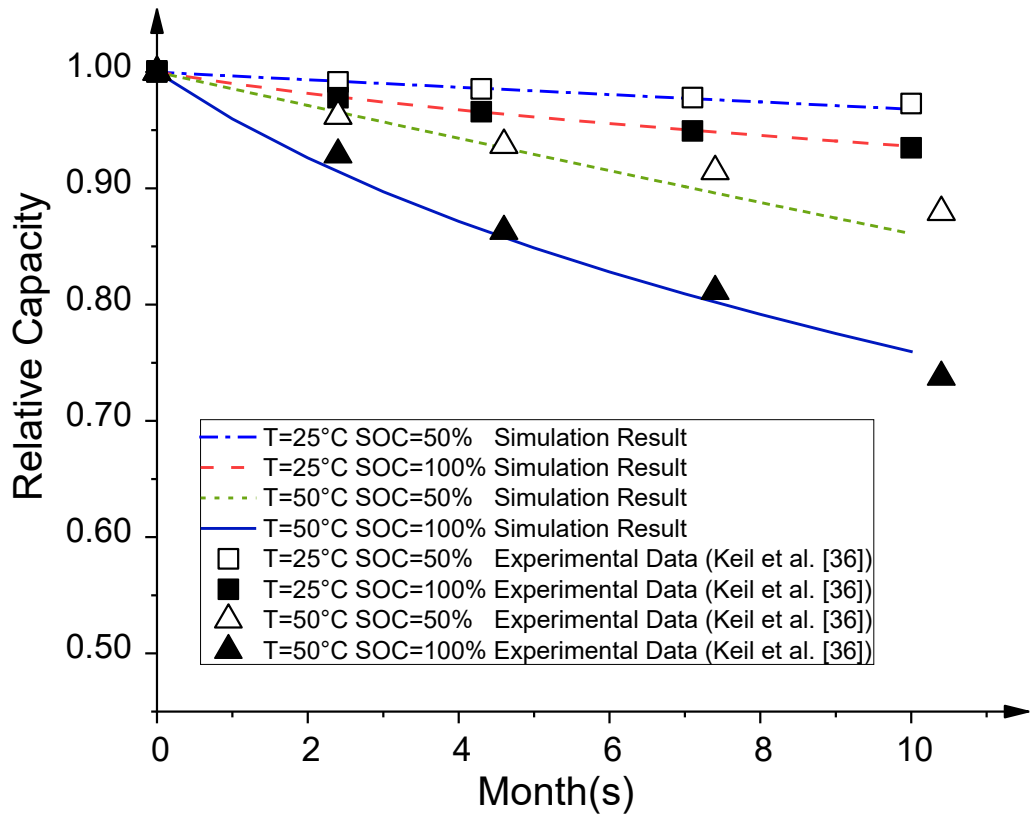

**Figure 6: Effects of SOC and cell working temperature on calendar loss of an NMC/graphite Li-ion battery cell. The experimental data is from [36].**

From Figure 6, it also can be seen that at 50 °C, the capacity drops nearly 24.0% of its original capacity after 10 months under 100% SOC, while the capacity drops around 14.0% under 50% SOC. The calendar capacity loss at 50 °C is higher than 25 °C. A larger calendar loss was observed at a high cell working temperature. The main reason is that the increased temperature increases the kinetic reaction for the side reaction; other minor reasons might be: at a higher temperature, a higher current passing through the SEI layer can be achieved since ionic conductivity of the SEI layer increase with the temperature, and consequently leading to a higher rate of the electrolyte decomposition at the surface of the SEI layer [30]. Moreover, Figure 7 shows the effect of particle size of the negative electrode on the calendar loss for the NMC-graphite Li-ion battery cell. It can be seen that a larger particle size ratio results in a smaller calendar loss. When the particle size ratio is 0.25, the capacity drops over 22% of its original capacity after 10 months. When the particle size ratio is 2, the capacity fade drops less than 2% of its original capacity. One possible reason is that as the particle size increases, the surface-to-volume ratio of the active material decreases for the side reactions, and consequently results in a smaller capacity loss [9,59].

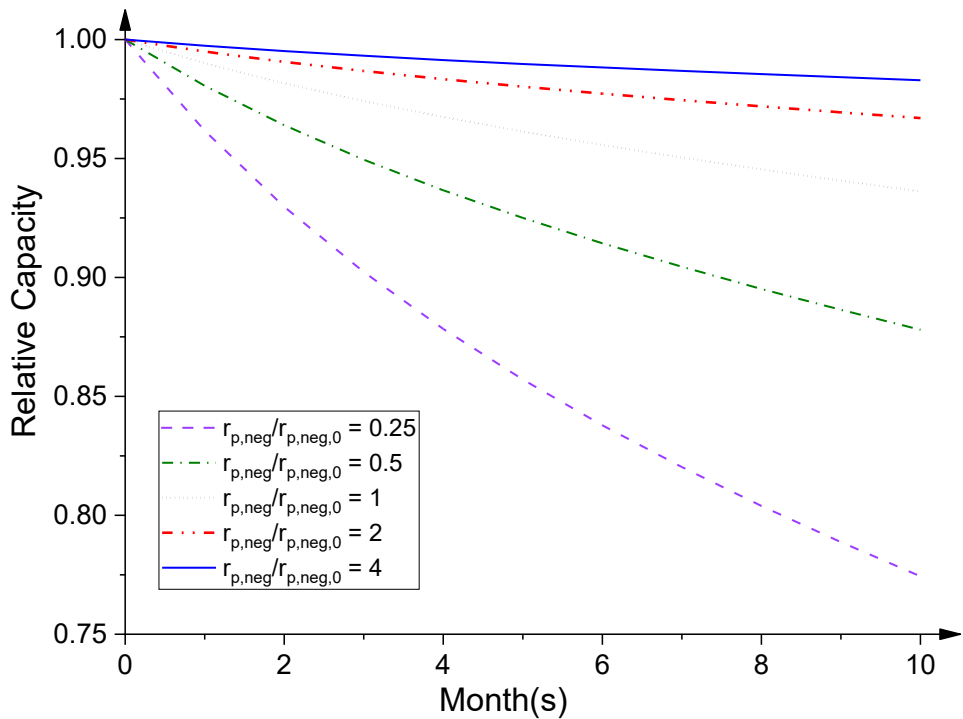

**Figure 7: Effect of negative particle size on calendar loss of the NMC/graphite Li-ion battery cell, where, $r_{p,neg}$ stands for the particle size on anode, $r_{p,neg,0}$= 26.2 μm represent the**

**reference particle size used in the model, coming from *Hosseinzadeh, et al.* [46], particles size at the positive side is fixed, cell working temperature: 25 °C and SOC: 100%).**

*3.3 Discussions*

In this model, the electrode kinetic was considered as a combination of the lithium ion intercalation and side reactions. However, some studies have considered that the "self-discharging" process is the outcome of side reactions at the anode side during the storage period, and thus the Li$^+$ intercalation can be ignored [29]. In fact, there could be some external electron leakages that lead to capacity loss [15]. In this work, the "self-discharging" process was treated as a combined effect of the intercalation and side reactions. In our simulation, under the condition of 100% SOC and 25 °C working temperature, it was found that the local current density of the anode side reaction is about 70% of the self-discharging current density. Therefore, it can be concluded that the anode side reaction is an important part of the "self-discharging", and the role of the intercalation in the "self-discharging" cannot be ignored. Since side reactions play such an important role in the "self-discharging" process, fitting $i_{0,side}$ based on the "self-discharging" property is still more reliable than using assumed value [29]. Our modeling suggests that the storage condition has a strong effect on battery lifetime. In practice, the effect of the storage condition on battery aging is an essential research area for electric vehicle (EV) batteries since their degradations can be mitigated through maintaining the battery cell temperature at an ideal condition. In this work, two different temperatures 25 °C and 50 °C were chosen to compare with the experimental data reported in the literature. For practical applications, such as EVs, the range of storage temperature is typically within the range of -10 °C to 40 °C [60]. Also, on one hand, we found that a larger electrode particle size leads to a smaller calendar capacity loss. On the other hand, a smaller electrode particle might lead to a larger battery capacity as observed in the experiments [61,62]. There might be a trade-off between the capacity loss and capacity output from the aspect of optimizing the electrode particle size. Other factors can also be affected by particle size, such as Coulombic efficiency [62]. Thus, optimizing the particle size might be another important aspect for achieving long-term and high-performance NMC-graphite Li-ion batteries that should be considered.

## Conclusions

To investigate the calendar capacity loss of NMC-graphite Li-ion batteries during the storage period, a physics-based model has been developed by incorporating electro-kinetics and side reactions to simulate the internal electrochemical processes under various storage conditions. Effects of state of charge (SOC), cell working temperature and negative electrode particle size on calendar loss of an NMC-graphite Li-ion battery under various storage conditions were examined. It was found that a smaller SOC, a lower cell working temperature and a larger particle size can decrease the calendar capacity loss and prolong the battery life of NMC-graphite Li-ion batteries. A trade-off between the calendar capacity loss and maximum capacity in terms of optimizing the electrode particle size is suggested to be considered in future work. This modeling work can provide certain useful guidance on how to minimize the calendar capacity loss and optimize the battery performance in future.

**CRediT authorship contribution statement**

**Boman Su:** Investigation, Methodology, Data curation, Formal analysis, Validation, Visualization, Writing-original draft; **Xinyou Ke:** Conceptualization, Methodology, Data curation, Formal analysis, Validation, Visualization, Writing-original draft; **Chris Yuan:** Funding acquisition, Project administration, Resources, Software, Supervision, Writing-original draft.

**Declaration of competing interest**

The authors declare that they have not known competing fanatical interests or personal relationships that could have appreared to influence the work reported in the paper.

**Acknowledgements**

This work was partially supported by Leonard Case Jr. Endowment Fund, the Critical Materials Institute, and the Industrial Assessment Center at Case Western Reserve University.

https://orbit.dtu.dk/en/activities/modelling-of-lithium-battery-and-v2g-charger-for-degradation-asse

15. Yazami, R., & Reynier, Y. F. (2002). Mechanism of self-discharge in graphite-lithium anode. *Electrochimica Acta*, *47*, 1217-1223.

16. Utsunomiya, T., Hatozaki, O., Yoshimoto, N., Egashira, M., & Morita, M. (2011). Self-discharge behavior and its temperature dependence of carbon electrodes in lithium-ion batteries. *Journal of Power Sources*, *196*, 8598-8603.

17. Byun, S., Park, J., Appiah, W. A., Ryou, M. H., & Lee, Y. M. (2017). The effects of humidity on the self-discharge properties of $LiNi_{1/3}Mn_{1/3}Co_{1/3}O_2$ /graphite and $LiCoO_2$/graphite lithium-ion batteries during storage. *RSC Advances*, *7*, 10915-10921.

18. Zilberman, I., Sturm, J., & Jossen, A. (2019). Reversible self-discharge and calendar aging of 18,650 nickel-rich, silicon-graphite lithium-ion cells. *Journal of Power Sources*, *425*, 217-226.

19. Redondo-Iglesias, E., Venet, P., & Pelissier, S. (2016, October). Measuring reversible and irreversible capacity losses on lithium-ion batteries. In 2016 *IEEE Vehicle Power and Propulsion Conference (VPPC)* (pp. 1-5), IEEE.

20. Arora, P., White, R. E., & Doyle, M. (1998). Capacity fade mechanisms and side reactions in lithium-ion batteries. *Journal of The Electrochemical Society*, *145*, 3647.

21. Zilberman, I., Ludwig, S., & Jossen, A. (2019). Cell-to-cell variation of calendar aging and reversible self-discharge in 18,650 nickel-rich, silicon–graphite lithium-ion cells. *Journal of Energy Storage*, *26*, 100900.

22. Lewerenz, M., Fuchs, G., Becker, L., & Sauer, D. U. (2018). Irreversible calendar aging and quantification of the reversible capacity loss caused by anode overhang. *Journal of Energy Storage*, *18*, 149-159.
30